
\input phyzzx.tex
\catcode`\@=11
\def\wlog#1{}
\def\eqname#1{\rel@x {\pr@tect
  \ifnum\equanumber<0 \xdef#1{{\rm\number-\equanumber}}%
     \gl@bal\advance\equanumber by -1
  \else \gl@bal\advance\equanumber by 1
     \ifx\chapterlabel\rel@x \def\d@t{}\else \def\d@t{.}\fi
    \xdef#1{{\rm\chapterlabel\d@t\number\equanumber}}\fi #1}}
\catcode`\@=12
\catcode`\@=11

\def\eat@#1{}
\mathchardef\prime@="0230
\def\prime{{{}\prime@{}}}
\def\prim@s{\prime@\futurelet\next\pr@m@s}

\def\,{\relax\ifmmode\mskip\thinmuskip\else\thinspace\fi}
\def\!{\relax\ifmmode\mskip-\thinmuskip\else\negthinspace\fi}
\def\frac#1#2{{#1\over#2}}
\def\dfrac#1#2{{\displaystyle{#1\over#2}}}

\def\:{\nobreak\hskip.1111em{:}\hskip.3333em plus .0555em\relax}
\def\intic@{\mathchoice{\hskip5\p@}{\hskip4\p@}{\hskip4\p@}{\hskip4\p@}}
\def\negintic@
 {\mathchoice{\hskip-5\p@}{\hskip-4\p@}{\hskip-4\p@}{\hskip-4\p@}}
\def\intkern@{\mathchoice{\!\!\!}{\!\!}{\!\!}{\!\!}}
\def\intdots@{\mathchoice{\cdots}{{\cdotp}\mkern1.5mu
    {\cdotp}\mkern1.5mu{\cdotp}}{{\cdotp}\mkern1mu{\cdotp}\mkern1mu
      {\cdotp}}{{\cdotp}\mkern1mu{\cdotp}\mkern1mu{\cdotp}}}
\newcount\intno@
\def\iint{\intno@=\tw@\futurelet\next\ints@}
\def\iiint{\intno@=\thr@@\futurelet\next\ints@}
\def\iiiint{\intno@=4 \futurelet\next\ints@}
\def\idotsint{\intno@=\z@\futurelet\next\ints@}
\def\ints@{\findlimits@\ints@@}
\newif\iflimtoken@
\newif\iflimits@
\def\findlimits@{\limtoken@false\limits@false\ifx\next\limits
 \limtoken@true\limits@true
   \else\ifx\next\nolimits\limtoken@true\limits@false
    \fi\fi}
\def\multintlimits@{\intop\ifnum\intno@=\z@\intdots@
  \else\intkern@\fi
    \ifnum\intno@>\tw@\intop\intkern@\fi
     \ifnum\intno@>\thr@@\intop\intkern@\fi\intop}
\def\multint@{\int\ifnum\intno@=\z@\intdots@\else\intkern@\fi
   \ifnum\intno@>\tw@\int\intkern@\fi
    \ifnum\intno@>\thr@@\int\intkern@\fi\int}
\def\ints@@{\iflimtoken@\def\ints@@@{\iflimits@
   \negintic@\mathop{\intic@\multintlimits@}\limits\else
    \multint@\nolimits\fi\eat@}\else
     \def\ints@@@{\multint@\nolimits}\fi\ints@@@}
\def\Sb{_\bgroup\vspace@
        \baselineskip=\fontdimen10 \scriptfont\tw@
        \advance\baselineskip by \fontdimen12 \scriptfont\tw@
        \lineskip=\thr@@\fontdimen8 \scriptfont\thr@@
        \lineskiplimit=\thr@@\fontdimen8 \scriptfont\thr@@
        \Let@\vbox\bgroup\halign\bgroup \hfil$\scriptstyle
            {##}$\hfil\cr}
\def\endSb{\crcr\egroup\egroup\egroup}
\def\Sp{^\bgroup\vspace@
        \baselineskip=\fontdimen10 \scriptfont\tw@
        \advance\baselineskip by \fontdimen12 \scriptfont\tw@
        \lineskip=\thr@@\fontdimen8 \scriptfont\thr@@
        \lineskiplimit=\thr@@\fontdimen8 \scriptfont\thr@@
        \Let@\vbox\bgroup\halign\bgroup \hfil$\scriptstyle
            {##}$\hfil\cr}
\def\endSp{\crcr\egroup\egroup\egroup}
\def\Let@{\relax\iffalse{\fi\let\\=\cr\iffalse}\fi}
\def\vspace@{\def\vspace##1{\noalign{\vskip##1 }}}
\def\aligned{\,\vcenter\bgroup\vspace@\Let@\openup\jot\m@th\ialign
  \bgroup \strut\hfil$\displaystyle{##}$&$\displaystyle{{}##}$\hfil\crcr}
\def\endaligned{\crcr\egroup\egroup}
\def\matrix{\,\vcenter\bgroup\Let@\vspace@
    \normalbaselines
  \m@th\ialign\bgroup\hfil$##$\hfil&&\quad\hfil$##$\hfil\crcr
    \mathstrut\crcr\noalign{\kern-\baselineskip}}
\def\endmatrix{\crcr\mathstrut\crcr\noalign{\kern-\baselineskip}\egroup
                \egroup\,}
\newtoks\hashtoks@
\hashtoks@={#}
\def\format{\crcr\egroup\iffalse{\fi\ifnum`}=0 \fi\format@}
\def\format@#1\\{\def\preamble@{#1}%
  \def\c{\hfil$\the\hashtoks@$\hfil}%
  \def\r{\hfil$\the\hashtoks@$}%
  \def\l{$\the\hashtoks@$\hfil}%
  \setbox\z@=\hbox{\xdef\Preamble@{\preamble@}}\ifnum`{=0 \fi\iffalse}\fi
   \ialign\bgroup\span\Preamble@\crcr}

\def\cases{\left\{\,\vcenter\bgroup\vspace@
     \normalbaselines\openup\jot\m@th
       \Let@\ialign\bgroup$##$\hfil&\quad$##$\hfil\crcr
      \mathstrut\crcr\noalign{\kern-\baselineskip}}

\newif\iftagsleft@
\tagsleft@true
\def\TagsOnRight{\global\tagsleft@false}
\def\tag#1$${\iftagsleft@\leqno\else\eqno\fi
 \hbox{\def\pagebreak{\global\postdisplaypenalty-\@M}%
 \def\nopagebreak{\global\postdisplaypenalty\@M}\rm(#1\unskip)}%
  $$\postdisplaypenalty\z@\ignorespaces}
\interdisplaylinepenalty=\@M
\def\allowdisplaybreak@{\def\allowdisplaybreak{\noalign{\allowbreak}}}
\def\displaybreak@{\def\displaybreak{\noalign{\break}}}
\def\align#1\endalign{\def\tag{&}\vspace@\allowdisplaybreak@\displaybreak@
  \iftagsleft@\lalign@#1\endalign\else
   \ralign@#1\endalign\fi}
\def\ralign@#1\endalign{\displ@y\Let@\tabskip\centering
   \halign to\displaywidth
     {\hfil$\displaystyle{##}$\tabskip=\z@&$\displaystyle{{}##}$\hfil
       \tabskip=\centering&\llap{\hbox{(\rm##\unskip)}}\tabskip\z@\crcr
             #1\crcr}}
\def\lalign@
 #1\endalign{\displ@y\Let@\tabskip\centering\halign to \displaywidth
   {\hfil$\displaystyle{##}$\tabskip=\z@&$\displaystyle{{}##}$\hfil
   \tabskip=\centering&\kern-\displaywidth
        \rlap{\hbox{(\rm##\unskip)}}\tabskip=\displaywidth\crcr
               #1\crcr}}
\def\overrightarrow{\mathpalette\overrightarrow@}
\def\overrightarrow@#1#2{\vbox{\ialign{$##$\cr
    #1{-}\mkern-6mu\cleaders\hbox{$#1\mkern-2mu{-}\mkern-2mu$}\hfill
     \mkern-6mu{\to}\cr
     \noalign{\kern -1\p@\nointerlineskip}
     \hfil#1#2\hfil\cr}}}
\def\overleftarrow{\mathpalette\overleftarrow@}
\def\overleftarrow@#1#2{\vbox{\ialign{$##$\cr
     #1{\leftarrow}\mkern-6mu\cleaders
      \hbox{$#1\mkern-2mu{-}\mkern-2mu$}\hfill
      \mkern-6mu{-}\cr
     \noalign{\kern -1\p@\nointerlineskip}
     \hfil#1#2\hfil\cr}}}
\def\overleftrightarrow{\mathpalette\overleftrightarrow@}
\def\overleftrightarrow@#1#2{\vbox{\ialign{$##$\cr
     #1{\leftarrow}\mkern-6mu\cleaders
       \hbox{$#1\mkern-2mu{-}\mkern-2mu$}\hfill
       \mkern-6mu{\to}\cr
    \noalign{\kern -1\p@\nointerlineskip}
      \hfil#1#2\hfil\cr}}}
\def\underrightarrow{\mathpalette\underrightarrow@}
\def\underrightarrow@#1#2{\vtop{\ialign{$##$\cr
    \hfil#1#2\hfil\cr
     \noalign{\kern -1\p@\nointerlineskip}
    #1{-}\mkern-6mu\cleaders\hbox{$#1\mkern-2mu{-}\mkern-2mu$}\hfill
     \mkern-6mu{\to}\cr}}}
\def\underleftarrow{\mathpalette\underleftarrow@}
\def\underleftarrow@#1#2{\vtop{\ialign{$##$\cr
     \hfil#1#2\hfil\cr
     \noalign{\kern -1\p@\nointerlineskip}
     #1{\leftarrow}\mkern-6mu\cleaders
      \hbox{$#1\mkern-2mu{-}\mkern-2mu$}\hfill
      \mkern-6mu{-}\cr}}}
\def\underleftrightarrow{\mathpalette\underleftrightarrow@}
\def\underleftrightarrow@#1#2{\vtop{\ialign{$##$\cr
      \hfil#1#2\hfil\cr
    \noalign{\kern -1\p@\nointerlineskip}
     #1{\leftarrow}\mkern-6mu\cleaders
       \hbox{$#1\mkern-2mu{-}\mkern-2mu$}\hfill
       \mkern-6mu{\to}\cr}}}
\def\sqrt#1{\radical"270370 {#1}}
\def\dots{\relax\ifmmode\let\next=\ldots\else\let\next=\tdots@\fi\next}
\def\tdots@{\unskip\ \tdots@@}
\def\tdots@@{\futurelet\next\tdots@@@}
\def\tdots@@@{$\mathinner{\ldotp\ldotp\ldotp}\,
   \ifx\next,$\else
   \ifx\next.\,$\else
   \ifx\next;\,$\else
   \ifx\next:\,$\else
   \ifx\next?\,$\else
   \ifx\next!\,$\else
   $ \fi\fi\fi\fi\fi\fi}
\def\text{\relax\ifmmode\let\next=\text@\else\let\next=\text@@\fi\next}
\def\text@@#1{\hbox{#1}}
\def\text@#1{\mathchoice
 {\hbox{\everymath{\displaystyle}\def\textfonti{\the\textfont1 }%
    \def\textfontii{\the\textfont2 }\textdef@@ T#1}}
 {\hbox{\everymath{\textstyle}\def\textfonti{\the\textfont1 }%
    \def\textfontii{\the\textfont2 }\textdef@@ T#1}}
 {\hbox{\everymath{\scriptstyle}\def\textfonti{\the\scriptfont1 }%
   \def\textfontii{\the\scriptfont2 }\textdef@@ S\rm#1}}
 {\hbox{\everymath{\scriptscriptstyle}%
   \def\textfonti{\the\scriptscriptfont1 }%
   \def\textfontii{\the\scriptscriptfont2 }\textdef@@ s\rm#1}}}
\def\textdef@@#1{\textdef@#1\rm \textdef@#1\bf
   \textdef@#1\sl \textdef@#1\it}

\def\textdef@#1#2{%
 \def\next{\csname\expandafter\eat@\string#2fam\endcsname}%
\if S#1\edef#2{\the\scriptfont\next\relax}%
 \else\if s#1\edef#2{\the\scriptscriptfont\next\relax}%
 \else\edef#2{\the\textfont\next\relax}\fi\fi}
\scriptfont\itfam=\tenit \scriptscriptfont\itfam=\tenit
\scriptfont\slfam=\tensl \scriptscriptfont\slfam=\tensl
\mathcode`\0="0030
\mathcode`\1="0031
\mathcode`\2="0032
\mathcode`\3="0033
\mathcode`\4="0034
\mathcode`\5="0035
\mathcode`\6="0036
\mathcode`\7="0037
\mathcode`\8="0038
\mathcode`\9="0039
\def\Cal{\relax\ifmmode\let\next=\Cal@\else
    \def\next{\errmessage{Use \string\Cal\space only in %
      math mode}}\fi\next}
    \def\Cal@#1{{\fam2 #1}}
\def\bold{\relax\ifmmode\let\next=\bold@\else
    \def\next{\errmessage{Use \string\bold\space only in %
      math mode}}\fi\next}
    \def\bold@#1{{\fam\bffam #1}}
\mathchardef\Gamma="0000
\mathchardef\Delta="0001
\mathchardef\Theta="0002
\mathchardef\Lambda="0003
\mathchardef\Xi="0004
\mathchardef\Pi="0005
\mathchardef\Sigma="0006
\mathchardef\Upsilon="0007
\mathchardef\Phi="0008
\mathchardef\Psi="0009
\mathchardef\Omega="000A
\mathchardef\varGamma="0100
\mathchardef\varDelta="0101
\mathchardef\varTheta="0102
\mathchardef\varLambda="0103
\mathchardef\varXi="0104
\mathchardef\varPi="0105
\mathchardef\varSigma="0106
\mathchardef\varUpsilon="0107
\mathchardef\varPhi="0108
\mathchardef\varPsi="0109
\mathchardef\varOmega="010A
\def\wlog#1{\immediate\write-1{#1}}
\catcode`\@=12  
\def\=def{\; \mathop{=}_{\text{\rm def}} \;}
\def\rd{\partial}
\def\Res{\mathop{\;\text{Res}\;}}

\def\bfZ{{\bold Z}}
\def\calA{{\cal A}}
\def\calB{{\cal B}}
\def\calL{{\cal L}}
\def\calM{{\cal M}}
\def\calO{{\cal O}}
\def\rdtilde{{\tilde{\rd}}}
\def\lambdatilde{{\tilde{\lambda}}}
\def\tauB{\tau^{\text{B}}}
\def\FB{F^{\text{B}}}
\def\WB{W^{\text{B}}}
\def\wB{w^{\text{B}}}
\def\ZB{Z^{\text{B}}}
\hsize=15.5truecm
\vsize=23truecm
\sequentialequations
\doublespace
\TagsOnRight
\overfullrule=0pt
\sectionstyle={\Number}
\pubnum={Kyoto University KUCP-0058/93}
\date={January 1993}
\titlepage
\title{\fourteencp Quasi-classical limit of BKP hierarchy
  and W-infinity symmetries}
\author{Kanehisa Takasaki}
\address{
  Department of Fundamental Sciences\break
  Faculty of Integrated Human Studies, Kyoto University\break
  Yoshida-Nihonmatsu-cho, Sakyo-ku, Kyoto 606, Japan\break
  E-mail: takasaki @ jpnyitp (Bitnet)\break
}
\abstract
\noindent
Previous results on quasi-classical limit of the KP and Toda
hierarchies are now extended to the BKP hierarchy. Basic
tools such as the Lax representation, the Baker-Akhiezer
function and the tau function are reformulated so as to fit
into the analysis of quasi-classical limit. Two subalgebras
$\WB_{1+\infty}$ and $\wB_{1+\infty}$ of the W-infinity
algebras $W_{1+\infty}$ and $w_{1+\infty}$ are introduced
as fundamental Lie algebras of the BKP hierarchy and its
quasi-classical limit, the dispersionless BKP hierarchy.
The quantum W-infinity algebra $\WB_{1+\infty}$ emerges in
symmetries of the BKP hierarchy. In quasi-classical limit,
these $\WB_{1+\infty}$ symmetries are shown to be contracted
into $\wB_{1+\infty}$ symmetries of the dispersionless BKP
hierarchy.

\endpage
\section{Introduction}

\noindent The notion of quasi-classical limit of integrable
hierarchies
[\REF\displess{
  Lebedev, D., and Manin, Yu.,
  Conservation laws and Lax representation on
  Benny's long wave equations,
  Phys.Lett. 74A (1979), 154--156. \nextline
  Kodama, Y.,
  A method for solving the dispersionless KP equation and
  its exact solutions,
  Phys. Lett. 129A (1988), 223-226;
  Solutions of the dispersionless Toda equation,
  Phys. Lett. 147A (1990), 477-482. \nextline
  Kodama, Y., and Gibbons, J.,
  A method for solving the dispersionless KP hierarchy and
  its exact solutions, II,
  Phys. Lett. 135A (1989), 167-170.}
\displess]
has been one of attractive topics in recent studies
of nonlinear integrable systems.  Of particular interest are
its relation to two dimensional quantum gravity and topological
conformal field theory
[\REF\TCFT{
  Dijkgraaf, R., Verlinde, H., and Verlinde, E.,
  Topological strings in $d<1$,
  Nucl. Phys. B352 (1991), 59-86.\nextline
  Krichever, I.M.,
  The dispersionless Lax equations and topological minimal models,
  Commun. Math. Phys. 143 (1991), 415-426. \nextline
  Dubrovin, B.A.,
  Hamiltonian formalism of Whitham-type hierarchies
  and topological Landau-Ginsburg models,
  Commun. Math. Phys. 145 (1992), 195-207.}
\TCFT]
and the underlying algebraic structure, so called W-infinity
algebras
[\REF\Winf{
  Bakas, I.,
   The structure of the $W_\infty$ algebra,
  Commun. Math. Phys. 134 (1990), 487-508.\nextline
  Park, Q-Han,
  Extended conformal symmetries in real heavens,
  Phys. Lett. 236B (1990), 429-432.\nextline
  Takasaki, K., and Takebe, T.,
  SDiff(2) Toda equation -- hierarchy,
  tau function and symmetries,
  Lett. Math. Phys. 23 (1991), 205-214;
  SDiff(2) KP hierarchy,
  in: {\it Infinite Analysis\/}, RIMS Research Project 1991,
  Int. J. Mod. Phys. A7, Suppl. 1B (1992), 889-922.}
\Winf].

This note is a sequel to recent papers
[\REF\Prev{
  Takasaki, K., and Takebe, T.,
  Quasi-classical limit of KP hierarchy,
  W-symmetries and free fermions,
  Kyoto preprint KUCP-0050/92 (July, 1992);
  Quasi-classical limit of Toda hierarchy
  and W-infinity symmetries,
  in preparation.}
\Prev]
on quasi-classical (or dispersionless) limit of the KP and Toda
hierarchies.  We here consider a KP hierarchy of B type, the BKP
hierarchy
[\REF\DJKM{
  Date, E., Kashiwara, M., Jimbo, M., and Miwa, T.,
  Transformation groups for soliton equations,
  in: {\it Nonlinear Integrable Systems ---
  Classical Theory and Quantum Theory}
  (World Scientific, 1983).}
\DJKM]
from the same point of view. The BKP hierarchy is a kind of
reduction of the ordinary KP hierarchy, but the machinery of
reduction is somewhat distinct from the KdV and generalized
KdV reductions.  Our main concern lies in the fate of W-infinity
symmetries of the KP and dispersionless KP hierarchies in the
course of reduction.

\section{Lax and dressing operators of BKP hierarchy}

\noindent Let $L$ be the Lax operator of the KP hierarchy,
$$
  L = \rd_x + \sum_{n=1}^\infty u_{n+1}(t) \rd_x^{-n}, \quad
  \rd_x = \rd/\rd x,                                 \tag\eq
$$
where the coefficients $u_n$ are unknown functions of a set of
flow variables $t$. Throught this note, the space variable $x$
is identified with the first flow variable
$$
  \quad x = t_1.                     \tag\eq
$$
Given a pseudo-differential operator $P$, let $P^*$ denote the
formal adjoint:
$$
  P = \sum p_n \rd_x^n \ \Rightarrow \
  P^* = \sum (-\rd_x)^n \cdot p_n.                       \tag\eq
$$
The BKP hierarchy is a reduction of the KP hierarchy by
the constraint [\DJKM]
$$
    \rd_x^{-1} L^* \rd_x = - L.                          \tag\eq
$$
This constraint is preserved only by ``odd" flows, $t_3,t_5,\ldots$,
of the KP hierarchy. These flows are given by the Lax equations
$$
  \dfrac{\rd L}{\rd t_{2n+1}} = [ B_{2n+1}, L ], \quad
  B_{2n+1} = \left( L^{2n+1} \right)_{\ge 0},            \tag\eq
$$
where, as usual, $(\quad)_{\ge 0}$ denotes the projection onto
nonnegative powers of $\rd_x$.

The above constraint can be translated into the language of a
dressing operator. As in the KP hierarchy, one can find a
dressing operator
$$
  W = 1 + \sum_{n=1}^\infty w_n(t) \rd_x^{-n}            \tag\eq
$$
that satisfies the dressing relation
$$
  L = W \rd_x W^{-1}                        \tag\eqname\Dress
$$
and the equations of flows
$$
  \dfrac{\rd W}{\rd t_{2n+1}} = B_{2n+1} W - W \rd_x^{2n+1}.
                                            \tag\eqname\FlowOfW
$$
In terms of $W$, the constraint to $L$ can rewritten
$$
  [ W^* \rd_x W, \rd_x ] = 0.                         \tag\eq
$$
This implies that $W^* \rd_x W$ is a pseudo-differential
operator with constant coefficients of the form
$$
  W^* \rd_x W = \rd_x + \text{(lower order terms)}    \tag\eq
$$
One can eliminate (or ``gauge away") the tail of lower order
terms by a residual gauge transformation
$$
  W \ \to \ W ( 1 + c_1 \rd_x^{-1} + c_2 \rd_x^{-2} + \cdots),
  \quad c_n = \text{const.},
                                                  \tag\eq
$$
which leaves (\Dress) and (\FlowOfW) invariant.  One can thus
single out a dressing operator that satisfies the condition
$$
  W^* \rd_x W = \rd_x.                       \tag\eqname\ConW
$$
This gives an ultimate form of the constraint that $W$ has to
satisfy.

Having this choice of $W$, we now define the second Lax operator
$$
\align
  M =& W \left( \sum_{n=1}^\infty (2n+1)t_{2n+1} \rd_x^{2n+1}
                + x \rd_x \right) W^{-1}                        \\
    =& \sum_{n=1}^\infty (2n+1) t_{2n+1} L^{2n+1} + x L
     + \sum_{n=1}^\infty v_{n+1}(t) L^{-n}.           \tag\eq   \\
\endalign
$$
By construction, $M$ satisfies the constraint
$$
  \rd_x^{-1} M^* \rd_x = - M.                          \tag\eq
$$
Furthermore, as in the KP hierarchy, $M$ satisfies the Lax equations
$$
  \dfrac{\rd M}{\rd t_{2n+1}} = [B_{2n+1}, M]          \tag\eq
$$
and the canonical commutation relation
$$
  [ L, M ] = L.                                        \tag\eq
$$
It should be noted that these $L$ and $M$ amount to
$L$ and $ML^{-1}$ in the KP hierarchy.

\section{Quasi-classical (or dispersionless) limit}

\noindent We now introduce a parameter $\hbar$ (``Planck constant")
and assume that the coefficients $u_n$ and $v_n$ of the Lax operators
depend on $\hbar$ as well as $t$ and behave smoothly as $\hbar \to 0$:
$$
\align
  u_n(\hbar,t) =& u^{(0)}_n(t) + O(\hbar),            \\
  v_n(\hbar,t) =& v^{(0)}_n(t) + O(\hbar).    \tag\eq \\
\endalign
$$
Furthermore, we modify the previous Lax formalism of the BKP
hierarchy by replacing
$$
  \rd_x \to \hbar \rd_x, \quad
  \dfrac{\rd}{\rd t_{2n+1}} \to \hbar \dfrac{\rd}{\rd t_{2n+1}}.
                                               \tag\eq
$$
The Lax operators are now written
$$
\align
  L =& \hbar\rd_x
       + \sum_{n=1}^\infty u_{n+1}(\hbar,t)(\hbar\rd_x)^{-n},     \\
  M =& \sum_{n=1}^\infty (2n+1) t_{2n+1} L^{2n+1} + x L
       + \sum_{n=1}^\infty v_{n+1}(\hbar,t) L^{-n},        \tag\eq\\
\endalign
$$
and the Lax equations and the canonical commutation relation
become
$$
  \hbar \dfrac{\rd L}{\rd t_{2n+1}} = [ B_{2n+1}, L ], \quad
  \hbar \dfrac{\rd M}{\rd t_{2n+1}} = [ B_{2n+1}, M ],
                                                \tag\eqname\Laxhbar
$$
and
$$
  [L, M ] = \hbar L                             \tag\eqname\CCRhbar
$$
respectively.

In quasi-classical ($\hbar \to 0$) limit, commutators of
pseudo-differential operators are replaced by Poisson bracket
by the rule
$$
  [\hbar \rd_x, x] = \hbar \ \to \ \{ k, x \} = 1,          \tag\eq
$$
where $k$ is a conjugate variable of $x$. Lax equations (\Laxhbar)
and canonical commutation relation (\CCRhbar) thereby turn into
the quasi-classical (or dispersionless) Lax equations
$$
  \dfrac{\rd \calL}{\rd t_{2n+1}} = \{ \calB_{2n+1}, \calL \}, \quad
  \dfrac{\rd \calM}{\rd t_{2n+1}} = \{ \calB_{2n+1}, \calM \},
$$
and the Poisson bracket relation
$$
  \{ \calL, \calM \} = \calL                                  \tag\eq
$$
of the Laurent series
$$
\align
  \calL =& k + \sum_{n=1}^\infty u^{(0)}_{n+1}(t) k^{-n},          \\
  \calM =& \sum_{n=1}^\infty (2n+1) t_{2n+1} \calL^{2n+1} + x \calL
         + \sum_{n=0}^\infty v^{(0)}_{2n+2}(t) \calL^{-2n-1},
                                               \tag\eqname\CalLCalM\\
\endalign
$$
where $\calB_n$ are given by
$$
  \calB_{2n+1} = \left( \calL^{2n+1} \right)_{\ge 0},    \tag\eq
$$
$(\quad)_{\ge 0}$ being the projection onto nonnegative powers
of $k$.  Similarly, the previous constraints to $L$ and $M$ are
replaced by constraints to $\calL$ and $\calM$:
$$
  \calL(t,-k) = - \calL(t,k), \quad
  \calM(t,-k) = - \calM(t,k).                            \tag\eq
$$
In other words, $\calL$ and $\calM$ are odd Laurent series of $k$;
this is a reason why the Laurent expansion of $\calM$ in (\CalLCalM)
contains only odd powers of $\calL$.  These $\calL$ and $\calM$,
like $L$ and $M$, correspond to $\calL$ and $\calM\calL^{-1}$ of
the dispersionless KP hierarchy.  Following the terminology in
the KP and Toda hierarchies [\displess], we call the above
hierarchy the ``dispersionless BKP hierarchy."

\section{Asymptotics of Baker-Akhiezer function and tau function}

\noindent
In the presence of $\hbar$, the Baker-Akhiezer function
$\Psi$ is given by the (formal) Laurent series
$$
\align
  & \Psi(\hbar,t,\lambda)
    = \left( 1 + \sum_{n=1}^\infty w_n(\hbar,t) \lambda^{-n} \right)
    \exp \hbar^{-1} t(\lambda),                                      \\
  & t(\lambda) = \sum_{n=0}^\infty t_{2n+1} \lambda^{2n+1}
               = \sum_{n=1}^\infty t_{2n+1} \lambda^{2n+1} + x \lambda,
                                                            \tag\eq  \\
\endalign
$$
of a parameter $\lambda$ (``spectral parameter").  By construction,
$\Psi$ satisfies the linear equations
$$
\align
  & \lambda \Psi = L \Psi, \quad
    \hbar \lambda \dfrac{\rd \Psi}{\rd \lambda} = M \Psi,          \\
  & \hbar \dfrac{\rd \Psi}{\rd t_{2n+1}} = B_{2n+1} \Psi.  \tag\eq \\
\endalign
$$
This implies that $\Psi$ has a WKB asymptotic form as $\hbar \to 0$:
$$
  \Psi = \exp[ \hbar^{-1} S(t,\lambda) + O(\hbar^0)].
                                              \tag\eqname\PsiWKB
$$
The phase function $S$ becomes an odd function of $\lambda$:
$$
  S(t,-\lambda) = - S(t,\lambda).                     \tag\eq
$$
As in the case of the KP and Toda hierarchies [\Prev],
the phase function satisfy a set of Hamilton-Jacobi equations,
and the Hamilton-Jacobi equations reproduce the previous Lax
formalism of the dispersionless BKP hierarchy by means of
a Legendre transformation. In particular, one can find a
direct relation between $S$ and $\calM$:
$$
  \lambda \dfrac{\rd S}{\rd \lambda} = \calM(\lambda)
  = \sum_{n=0}^\infty (2n+1) t_{2n+1} \lambda^{2n+1}
   +\sum_{n=0}^\infty v^{(0)}_{2n+2} \lambda^{-2n-1}.   \tag\eq
$$

The tau function $\tauB$ of the BKP hierarchy is known to coincide
with the square root of the KP tau function $\tau$ [\DJKM].
In the present setting, $\tauB$ is a function that satisfies the
relation
$$
\align
  & \Psi
    = \dfrac{ \exp[ -2 \hbar \rdtilde(\lambda^{-1})] \tauB(\hbar,t) }
            { \tauB(\hbar,t) }
      \exp \hbar^{-1} t(\lambda),                                 \\
  & \rdtilde(\lambda^{-1})
    = \sum_{n=0}^\infty \dfrac{\lambda^{-2n-1}}{2n+1}
                        \dfrac{\rd}{\rd t_{2n+1}}.
                                                          \tag\eq  \\
\endalign
$$
To be consistent with WKB asymptotic form (\PsiWKB) of the
Baker-Akhiezer function in quasi-classical ($\hbar \to 0$)
limit, the tau function turns out to behave as
$$
   \tauB(\hbar,t) = \exp[ \hbar^{-2} \FB(t) + O(\hbar^{-1})]
                                            \tag\eqname\tauQC
$$
with a suitable scaling function $\FB(t)$ (``free energy").
Since $\tauB = \sqrt{\tau}$, the free energy is related to
the KP free energy $F$ as $\FB = F/2$. The phase function
$S$ can now be written
$$
  S(t,\lambda)  = t(\lambda) - 2\sum_{n=0}^\infty
                      \dfrac{\lambda^{-2n-1}}{2n+1}
                      \dfrac{\rd \FB}{\rd t_{2n+1}}.  \tag\eq
$$

\section{W-infinity symmetries in terms of Lax and dressing operators}

\noindent Let us first consider W-infinity symmetries of the BKP
hierarchy, putting $\hbar = 1$.  Following the case of the KP
hierarchy
[\REF\Orlov{
  Orlov, A.Yu., and Schulman, E.I.,
  Additional symmetries for integrable equations
  and conformal algebra representation,
  Lett. Math. Phys. 12 (1986) 171-179.\nextline
  Orlov, A.Yu.,
  Vertex operators, $\bar{\partial}$-problems,
  symmetries, variational indentities and
  Hamiltonian formalism for $2+1$ integrable systems,
  in: {\it Plasma Theory and Nonlinear and
  Turbulent Processes in Physics\/}
  (World Scientific, Singapore, 1988).\nextline
  Grinevich, P.G., and Orlov, A.Yu.,
  Virasoro action on Riemann surfaces, Grassmannians,
  $\det\bar{\partial}_j$ and Segal Wilson $\tau$ function,
  in: {\it Problems of modern quantum field theory\/}
  (Springer-Verlag, 1989).}
\Orlov],
we seek to construct W-infinity symmetries in such a form as
$$
  \delta_A W = A(L,M)_{\le -1} W,                     \tag\eq
$$
where $A(L,M)$, the data of a symmetry, is a non-commutative
Laurent series of $L$ and $M$,
$$
  A(L,M) = \sum_{i\in\bfZ,j\ge0}  a_{ij} L^i M^j,     \tag\eq
$$
and $(\quad)_{\le -1}$ the projection onto negative powers of
$\rd_x$. If $A$ is arbitrary, this gives a generic $W_{1+\infty}$
symmetry of the KP hierarchy.  The Lax operators are transformed
as
$$
  \delta_A L = [ A(L,M)_{\le -1}, L ], \quad
  \delta_A M = [ A(L,M)_{\le -1}, M ].                 \tag\eq
$$

We now have to find a condition under which $\delta_A$ preserves
the constraint to $W$ to the effect that
$$
  \rd_x^{-1} W^* \rd_x = W^{-1} \ \Rightarrow \
  \delta_A( \rd_x^{-1} W^* \rd_x) = \delta_A(W^{-1}).  \tag\eq
$$
Here $\delta_A$, by definition, is understood to act on both sides
of the last relation as:
$$
\align
  \delta_A( \rd_x^{-1} W^* \rd_x)
    =& \rd_x^{-1} \cdot (\delta_A W)^* \cdot \rd_x,            \\
  \delta_A( W^{-1})
    =& - W^{-1} \cdot \delta_A W \cdot W^{-1}.       \tag\eq \\
\endalign
$$
One can then prove, after somewhat lengthy technical calculations
of pseudo-differential operators, that $\delta_A$ fulfills the
above condition if
$$
   \rd_x^{-1} A(L,M)^* \rd_x = - A(L,M).         \tag\eqname\CondA
$$
Actually, this condition is equivalent to
$$
   \rd_x^{-1} A(\rd_x, x\rd_x)^* \rd_x = - A(\rd_x, x\rd_x).
                                                 \tag\eq
$$

Pseudo-differential operators satisfying the above condition
form a Lie subalgebra $\WB_{1+\infty}$ of $W_{1+\infty}$.
The W-infinity algebra $W_{1+\infty}$ is now realized by general
pseudo-differential operators. The Lie algebra homomorphism
$A \to - \rd_x^{-1} A^* \rd_x$ of this Lie algebra into itself
is obviously an involution.  The positive eigenspace,
$\WB_{1+\infty}$, thus becomes a Lie subalgebra:
$$
  W_{1+\infty} \supset \WB_{1+\infty}
  = \{ A(\rd_x,x\rd_x) \mid \rd_x^{-1} A^* \rd_x = -A \}.  \tag\eq
$$
The Lax operators $L$ and $M$ are elements of this Lie subalgebra.
The dressing operator $W$ may be thought of as an element of an
associated ``W-infinity group." Such a group actually does not
exist in a mathematically rigorous sense, but may be realized
as a kind of formal group whose generic elements are written
$g = \exp \epsilon A$, where $A$ lies in $\WB_{1+\infty}$ and
$\epsilon$ is a formal parameter, and satisfy the condition
$$
  \rd_x^{-1} g^* \rd_x = g^{-1}.                           \tag\eq
$$

These $\WB_{1+\infty}$ symmetries give rise to symmetries of
the dispersionless BKP hierarchy as follows. Let $\hbar$
now take a generic value.  As $\hbar \to 0$, the Lax operators
$L$ and $M$ are replaced by their quasi-classical counterparts
$\calL$ and $\calM$, and we are left with symmetries of the
dispersionless BKP hierarchy of the form
$$
  \delta_\calA \calL = \{ \calA(\calL,\calM)_{\le -1}, \calL \},
  \quad
  \delta_\calA \calM = \{ \calA(\calL,\calM)_{\le -1}, \calM \},
                                                          \tag\eq
$$
where $\calA$ is now a Laurent series of $\calL$ and $\calM$,
$$
  \calA(\calL,\calM) = \sum_{i\in\bfZ,j\ge 0} a_{ij} \calL^i \calM^j,
                                                    \tag\eq
$$
and $(\quad)_{\le -1}$ the projection onto negative powers of $k$.
In place of (\CondA), $\calA$ has to satisfy the condition
$$
  \calA(-\lambda,-\mu) = - \calA(\lambda,\mu).
                                            \tag\eqname\CondCalA
$$
This defines a Lie subalgebra $\wB_{1+\infty}$ of the classical
W-infinity algebra $w_{1+\infty}$,
$$
  w_{1+\infty} \supset \wB_{1+\infty}
  = \{ \calA(\lambda,\mu) \mid
       \calA(-\lambda,-\mu) = - \calA(\lambda,\mu) \},  \tag\eq
$$
with respect to the Poisson bracket
$$
  \{ \calA, \calB \} = \lambda \frac{\rd \calA}{\rd \lambda}
                       \frac{\rd \calB}{\rd \mu}
                     - \frac{\rd \calA}{\rd \mu}
                       \lambda \frac{\rd \calB}{\rd \lambda}.
                                                         \tag\eq
$$

\section{W-infinity symmetries in terms of tau function}

\noindent  The above W-infinity symmetries can also be derived
from a vertex operator.  The vertex operator for the BKP
hierarchy ($\hbar = 1$) discovered by Date et al. [\DJKM]
can be written
$$
  \ZB(\lambdatilde,\lambda)
  =  \frac{1}{2} \;
     \frac{\lambdatilde - \lambda}{\lambdatilde + \lambda}
     \left( \exp[ t(\lambdatilde) + t(\lambda)]
           \exp[ -2\rdtilde(\lambdatilde^{-1})
                 -2\rdtilde(\lambda^{-1}) ]
     - 1 \right).                                    \tag\eq
$$
This gives a two-parameter family of infinitesimal transformation
of the BKB tau function:
$\tauB \to \tauB + \epsilon \ZB(\lambdatilde,\lambda) \tauB$.
Furthermore, by a boson-fermion correspondence,
$\ZB(\lambdatilde,\lambda)$ corresponds to the fermion bilocal
operator $:\phi(\lambdatilde)\phi(\lambda):$ of a newtral
fermion field $\phi(\lambda)$ [\DJKM]. The previous
$\WB_{1+\infty}$ symmetry $\delta_A$ can be identified with
$$
  W_A = \oint A(\lambdatilde \frac{\rd}{\rd \lambdatilde},
              \lambdatilde) \ZB(\lambdatilde,\lambda)
              |_{\lambdatilde=-\lambda}
              \frac{d \log\lambda}{2 \pi i}          \tag\eq
$$
in the bosonic language, and with
$$
  \calO_A = \oint : A(\lambdatilde \frac{\rd}{\rd \lambdatilde},
                    \lambdatilde) \phi(\lambdatilde) \phi(\lambda)
                    |_{\lambdatilde=-\lambda} :
                  \frac{d \log\lambda}{2 \pi i}      \tag\eq
$$
in the fermionic language. If $\hbar$ takes a generic value,
one has to replace
$$
  t_{2n+1} \to \hbar^{-1} t_{2n+1},
  \quad
  \dfrac{\rd}{\rd t_{2n+1}} \to \hbar \dfrac{\rd}{\rd t_{2n+1}},
  \quad
  \lambdatilde\dfrac{\rd}{\rd \lambdatilde} \to
    \hbar \lambdatilde\dfrac{\rd}{\rd \lambdatilde}   \tag\eq
$$
in the above construction. Let $W_A(\hbar)$ and $\calO_A(\hbar)$
denote the corresponding symmetry generators.

We now show how these $\WB_{1+\infty}$ symmetries of the tau
function $\tauB$ can be reduced to $\wB_{1+\infty}$ symmetries
of the free energy $\FB$. To this end, we consider the special
generators
$$
  A_{ij} = \half ( L^i M^j + (-1)^{i+j} M^j L^i )     \tag\eq
$$
of $\WB_{1+\infty}$ and the corresponding bosonic symmetry
operators $\WB_{ij}(\hbar) = \WB_{A_{ij}}(\hbar)$. Recalling
asymptotic behavior (\tauQC) of the tau function, one can
easily calculate the action by $\WB_{ij}(\hbar)$ to the
lowest order of $\hbar$-expansion as:
$$
  \dfrac{ \WB_{ij}(\hbar) \tauB(\hbar,t) }
        { \tauB(\hbar,t) }
  = - \hbar^{-j-1} \left(  \frac{1}{j+1}
           \Res \lambda^i \calM(\lambda)^{j+1} d\log\lambda
           + O(\hbar) \right),                        \tag\eq
$$
where ``Res" means the formal residue,
$$
  \Res \lambda^n d\log\lambda = \delta_{n,0}.         \tag\eq
$$
Picking out the most singular term ($\propto \hbar^{-j-1}$),
we define
$$
\align
  \wB_{ij} \FB
  =& - \frac{1}{j+1} \Res \lambda^i \calM(\lambda)^{j+1} d\log\lambda \\
  =& - \frac{1}{j+1} \Res \calL^i \calM^{j+1} d\log\calL.
                                                       \tag\eq        \\
\endalign
$$
This gives a $\wB_{1+\infty}$ symmetry of the dispersionless
BKP hierarchy associated with the element
$$
  \calA_{ij} = \lambda^i \mu^j                         \tag\eq
$$
of $\wB_{1+\infty}$. The symmetry is indeed realized as the
infinitesimal transformation
$\FB \to \FB + \epsilon \wB_{ij}\FB$
of the free energy.

Note that $\wB_{ij}F$ vanishes if $i+j$ is an even number. Thus
only half of the above $\WB_{1+\infty}$ generators $A_{ij}$ (i.e.,
those with $i+j$ being odd) correspond to nontrivial $\wB_{1+\infty}$
symmetries of the BKP hierarchy. This is quite natural, because
$\calA_{ij}$ belong to $\wB_{1+\infty}$ if and only if $i+j$ is
odd.

\section{Conclusion}

\noindent Our previous results on quasi-classical limit of the
KP and Toda hierarchies can thus be extended to the BKP hierarchy.
We have been able to identify the two W-infinity algebras
$\WB_{1+\infty}$ and $\wB_{1+\infty}$ as fundamental Lie algebras
of the BKP and dispersionless BKP hierarchies.  These W-infinity
algebras indeed emerge in both the Lax formalism of the hierarchies
and the construction of symmetries.  As in the case of the KP and
Toda hierarchies, we have seen that these structures of the quantum
W-infinity algebra $\WB_{1+\infty}$ are smoothly contracted to
those of the classical W-infinity algebra $\wB_{1+\infty}$.

The author is very grateful to Takashi Takebe for a lot of fruitful
discussions. This work is supported in part by the Grant-in-Aid
for Scientific Researches, the Ministry of Education, Science and
Culture, Japan.

\refout
\bye